\documentclass[aps,prl,twocolumn,amsmath,twoside,floatfix,superscriptaddress,preprintnumbers]{revtex4}
\usepackage{graphicx}
\newcommand{\bea}{\begin{eqnarray}}
\newcommand{\beq}{\begin{equation}}
\newcommand{\eea}{\end{eqnarray}}
\newcommand{\eeq}{\end{equation}}

\newcommand{\lsim}{\raise0.3ex\hbox{$\;<$\kern-0.75em\raise-1.1ex\hbox{$\sim\;$}}}
\newcommand{\gsim}{\raise0.3ex\hbox{$\;>$\kern-0.75em\raise-1.1ex\hbox{$\sim\;$}}}
\newcommand{\bpm}{\begin{pmatrix}}
\newcommand{\epm}{\end{pmatrix}}

\newcommand{\unity}{{\hbox{1\kern-.8mm l}}}

\begin{document}

\title{Probing New Physics through $\mu-e$ Universality in $K\rightarrow \ell\nu$}

\author{A. Masiero}
\affiliation{ Dip. di Fisica `G. Galilei', Univ. di Padova and 
INFN, Sezione di Padova, Via Marzolo 8, I-35131, Padua, Italy}
\author{P. Paradisi}
\affiliation{Dip. di Fisica, Universit\`a di Roma II
``Tor Vergata'' and INFN, Sezione di Roma II,
Via della Ricerca Scientifica 1, I-00133, Rome, Italy}
\author{R. Petronzio}
\affiliation{Dip. di Fisica, Universit\`a di Roma II
``Tor Vergata'' and INFN, Sezione di Roma II,
Via della Ricerca Scientifica 1, I-00133, Rome, Italy}

\begin{abstract}

The recent NA48/2 improvement on 
$R_{K}=\Gamma(K\rightarrow e\nu)/\Gamma(K\rightarrow \mu\nu)$ 
emphasizes the role of $K_{l2}$ decays in probing the  $\mu-e$ universality.
Supersymmetric (SUSY) extensions of the Standard Model can exhibit
$\mu-e$ non-universal contributions. Their origin is twofold:
 those deriving from lepton flavor conserving couplings are subdominant 
with respect to those arising from lepton flavor violating (LFV) sources. 
We show that  $\mu-e$ non-universality  in $K_{\ell2}$
is quite effective in constraining relevant regions of SUSY models with LFV 
(for instance, supergravities with a see-saw mechanism for neutrino masses).
A comparison with analogous bounds coming from  $\tau$ LFV decays proves 
the relevance of the measurement of $R_{K}$ to probe LFV in SUSY.

\end{abstract}

\maketitle

\section{Introduction}
%\label{sect-intro}
%\textbf{Introduction:}
High precision electroweak tests represent a powerful tool to probe 
the Standard Model (SM) and, hence, to constrain or obtain indirect hints 
of new physics beyond it. Kaon and pion physics are 
obvious grounds where to perform such tests, for instance in the  
well studied  $\pi_{\ell2}$ ($\pi\rightarrow \ell\nu_{\ell}$) and 
$K_{\ell2}$ ($K\rightarrow \ell\nu_{\ell}$) decays, where $l= e$ or $\mu$. 
Unfortunately, the relevance of these single  decay channels in probing 
the SM is severely hindered by our theoretical uncertainties, 
which still remain at the percent level
(in particular due to the uncertainties on non perturbative quantities like
$f_{\pi}$ and $f_{K}$). This is what prevents us from fully exploiting 
such decay modes in constraining new physics, 
in spite of the fact that it is possible to obtain non-SM
contributions which exceed the high experimental precision which has been 
achieved on those modes.

On the other hand, in the ratios $R_{\pi}$ and $R_{K}$ of the electronic
and muonic decay modes
$R_{\pi}\!=\!\Gamma(\pi\!\rightarrow\!e\nu)/\Gamma(\pi\!\rightarrow\! \mu\nu)$
and $R_{K}\!=\!\Gamma(K\!\rightarrow \!e\nu)/\Gamma(K\!\rightarrow\! \mu\nu)$,
the hadronic uncertainties cancel to a very large extent.
As a result, the SM predictions of $R_{\pi}$ and $R_{K}$ are known 
with excellent accuracy \cite{f} and this makes it possible to fully 
exploit the great experimental resolutions on $R_{\pi}$ \cite{pdg} 
and $R_{K}$ \cite{pdg,na} to constrain new physics effects. 
Given our limited predictive power on  $f_{\pi}$ and $f_{K}$, deviations
from the $\mu-e$ universality represent the best hope we have at the moment
to detect new physics effects in $\pi_{\ell2}$ and $K_{\ell2}$.

The most recent NA48/2 result on $R_K$:
$$
R^{exp.}_{K}=(2.416\pm 0.043_{stat.} \pm 0.024_{syst.})
\cdot 10^{-5}\,\,\,\,\,\,\rm{NA48/2}
$$

which will further improve with current analysis, significantly improves on 
the previous PDG value:
$$
R^{exp.}_{K}=(2.44\pm 0.11)\cdot 10^{-5}.
$$
This is to be compared with the SM prediction which reads:
$$
R^{SM}_{K}=(2.472\pm 0.001)\cdot 10^{-5}.
$$
Denoting by $\Delta r^{e-\mu}_{\!NP}$ the deviation from $\mu-e$ 
universality in $R_{K}$ due to new physics, i.e.:
\beq
\label{one}
R_{K}=\frac{\Gamma^{K\rightarrow e\nu_e}_{SM}}
{\Gamma^{K\rightarrow \mu\nu_\mu}_{SM}}
\left(1+\Delta r^{e-\mu}_{\!NP}\right),
\eeq
the NA48/2 result requires (at the $2\sigma$ level): 
$$
-0.063\leq\Delta r^{e-\mu}_{\!NP}\leq 0.017 \,\,\,\,\,\,\rm{NA48/2}.
$$ 

In this Letter we consider low-energy minimal SUSY extensions 
of the SM (MSSM) with R parity as the source of new physics to be tested by $R_K$
\cite{giudiceRP}. The question we intend to address is whether
SUSY can cause deviations from $\mu-e$ universality in $K_{l2}$
at a level which can be probed with the present attained experimental
sensitivity, namely at the percent level.
We will show that i) it is indeed possible for regions of the
MSSM to obtain $\Delta r^{e-\mu}_{\!NP}$ of
$\mathcal{O}(10^{-2})$ and ii) such large contributions to
$K_{\ell2}$ do not arise from SUSY lepton flavor conserving (LFC) effects, but,
rather, from LFV ones.

At first sight, this latter statement may seem rather puzzling.
The $K\!\rightarrow \!e\nu_e$ and $K\!\rightarrow\! \mu\nu_{\mu}$ decays are LFC
and one could expect that it is through LFC SUSY contributions affecting
differently the  two decays that one obtains 
the dominant source  of lepton flavor non-universality in SUSY.
On the other hand, one can easily guess that, whenever new physics
intervenes in $K\!\rightarrow\! e\nu_e$ and $K\!\rightarrow\! \mu\nu_{\mu}$
to create a departure from the strict SM $\mu-e$ universality,
these new contributions will be proportional to the lepton masses;
hence, it may happen (and, indeed, this is what occurs in the SUSY case)
that LFC contributions are suppressed with respect to the LFV ones
by higher powers of the first two generations lepton masses
(it turns out that the first contributions to $\Delta r^{e-\mu}_{\!NP}$
from LFC terms arise at the cubic order in $m_{\ell}$, with $\ell=e,\mu$).
A second, important reason for such result is that among the LFV
contributions to $R_K$ one can select those which involve flavor changes
from the first two lepton generations to the third one with
the possibility of picking up terms proportional to the tau-Yukawa coupling
which can be large in the large $\rm{\tan\beta}$ regime
(the parameter $\rm{\tan\beta}$ denotes the ratio of Higgs vacuum expectation
values responsible for the up- and down- quark masses, respectively).
Moreover,  the relevant one-loop induced LFV Yukawa interactions are known 
\cite{bkl} to acquire an additional $\rm{\tan\beta}$ factor with respect 
to the tree level LFC Yukawa terms.
Thus, the loop suppression factor can be (partially) compensated in the 
large $\rm{\tan\beta}$ regime.

Finally, given the NA48/2 $R_K$ central value below the SM prediction, 
one may wonder whether SUSY contributions could have the correct sign 
to account for such an effect.
Although the above mentioned  LFV terms can only add positive contributions 
to $R_K$ (since their amplitudes cannot interfere with the SM one), 
it turns out that there exist LFC contributions arising from double LFV 
mass insertions (MI) in the scalar lepton propagators which can destructively 
interfere with the SM contribution.
We will show that there exist regions of the SUSY parameter space where 
the total $R_K$ arising from all such SM and SUSY terms is indeed lower 
than $R^{SM}_K$.
\section{$\mu-e$ universality in $\pi\rightarrow \ell\nu$ and 
$K\rightarrow \ell\nu$ decays}
%\textbf{$\mu-e$ universality in $\pi\rightarrow \ell\nu$ and $K\rightarrow \ell\nu$ decays:}
Due to the V-A structure of the weak interactions,
the SM contributions to $\pi_{\ell2}$ and $K_{\ell2}$ are helicity suppressed; 
hence, these processes are very sensitive to non-SM effects 
(such as multi-Higgs effects) which might induce an effective pseudoscalar 
hadronic weak current.

In particular, charged Higgs bosons ($H^\pm$) appearing in any model with 
two Higgs doublets (including the SUSY case) can contribute at tree level to 
the above processes inducing the following effects \cite{hou}:
\beq
\label{tree}
\frac{\Gamma(M\!\rightarrow\! \ell\nu)}{\Gamma_{\!SM}(M\!\rightarrow\! \ell\nu)}=
\left[1-\!\tan^{2}\!\beta
\left(\frac{m_{s,d}}{m_u\!+\!m_{s,d}}\right)\frac{m^{2}_{M}}{m^{2}_{H}}\right]^2
%\label{one}
\eeq
where $m_{u}$ is the mass of the up quark while $m_{s,d}$ stands for the
down-type quark mass of the $M$ meson ($M=K, \pi$).
From Eq.~(\ref{tree}) it is evident that such tree level contributions do not 
introduce any lepton flavour dependent correction.
The first SUSY contributions violating the $\mu-e$ universality in 
$\pi\rightarrow \ell\nu$ and $K\rightarrow \ell\nu$ decays
arise at the one-loop level with various diagrams involving exchanges of
(charged and neutral) Higgs scalars, charginos, neutralinos and sleptons.
For our purpose, it is relevant to divide
all such contributions into two classes:
i) LFC contributions where the charged meson M decays without FCNC 
in the leptonic sector, i.e. $M\rightarrow \ell\nu_{\ell}$;
ii) LFV contributions $M\rightarrow \ell_i\nu_k$, with $i$ and $k$ 
referring to different generations (in particular, the interesting 
case will be for $i= e,\mu$, and $k=\tau$).

\section{The lepton flavour conserving case}
%\textbf{The lepton flavour conserving case:}
One-loop  corrections to $R_{\pi}$ and $R_{K}$ include box, wave function 
renormalization and vertex contributions from SUSY particle exchange.
The complete calculation of the $\mu$ decay in the MSSM \cite{pok} 
can be easily applied to the meson decays.
It turns out that all these LFC contributions yield values of
$\Delta r^{e-\mu}_{\!K\,SUSY}$ which are much smaller than the percent
level required by the achieved experimental sensitivity.
Indeed,  a typical $\Delta r^{e-\mu}_{SUSY}$ induced by (charged and neutral)
Higgs exchanges is of order
\beq
\label{lfc1}
\Delta r^{e-\mu}_{SUSY}\sim \frac{\alpha_{2}}{4\pi}
\left(\!\frac{m^{2}_{\mu}-m^{2}_{e}}{m^{2}_{H}}\!\right)\tan^2\beta\,,
\eeq
where  $H$ denotes a heavy Higgs circulating in the loop.
Then, even if we assume particularly favorable circumstances like
$\tan\beta=50$ and arbitrary relations among the Higgs boson masses,
we end up with  $\Delta r^{e-\mu}_{SUSY}\leq 10^{-6}$
much below the percent level of experimental sensitivity.

The charginos/neutralinos sleptons ($\tilde \ell_{e,\mu}$)
contributions to $\Delta r^{e-\mu}_{SUSY}$ are of the form
\beq
\label{lfc2}
\Delta r^{e-\mu}_{SUSY}\sim \frac{\alpha_{2}}{4\pi}
\left(\frac{\tilde{m}^{2}_{\mu}-\tilde{m}^{2}_{e}}
{\tilde{m}^{2}_{\mu}+\tilde{m}^{2}_{e}}\right)
\frac{m^{2}_{W}}{M^{2}_{SUSY}},
\eeq
where we considered all SUSY masses involved in the loops to be of 
$\mathcal{O}(M_{SUSY})$. The degeneracy of slepton masses 
(in particular those of the first two generations) 
severely suppresses these contributions.
Even if we assume a quite large mass splitting among slepton masses 
(at the $10\%$ level for instance) we end up with 
$\Delta r^{e-\mu}_{SUSY}\leq 10^{-4}$.
For the box-type non-universal contributions we find similar or even 
more suppressed effects compared to those we have studied.

On the other hand, one could wonder whether the quantity 
$\Delta r^{e-\mu}_{\!\,SUSY}$ can be constrained by the pion physics. 
In principle, the sensitivity could be even higher: from 
$$
R^{exp.}_{\pi}=
(1.230\pm 0.004)\cdot 10^{-4}\,\,\,\,\,\,\,\rm{PDG}
$$ 
and by making a comparison with the SM prediction
$$
R^{SM}_{\pi}=(1.2354\pm 0.0002)\cdot 10^{-4}
$$
one obtains (at the $2\sigma$ level) 
$$
-0.0107\leq\Delta r^{e-\mu}_{\!NP}\leq 0.0022.
$$
Unfortunately, even in the most favorable cases, 
$\Delta r^{e-\mu}_{\!\,SUSY}$ remains much below its actual exp. upper bound.

In conclusion, SUSY effects  with flavor conservation in the leptonic sector 
can differently contribute to the $K\rightarrow e\nu_e$ and 
$K\rightarrow \mu\nu_{\mu}$ decays, hence inducing a $\mu-e$ non-universality 
in $R_K$, however such effects are still orders of magnitude below 
the level of the present exp. sensitivity  on $R_K$.
The same conclusions hold for $R_{\pi}$.

\section{The lepton flavour violating case}
%\textbf{The lepton flavour violating case:}
It is well known that models containing at least two Higgs doublets 
generally allow flavour violating couplings of the Higgs bosons with 
the fermions \cite{bkq}. In the MSSM such LFV couplings are absent at 
tree level. However, once non holomorphic terms are generated by 
loop effects (so called HRS corrections \cite{hrs}) and given a source 
of LFV among the sleptons, Higgs-mediated (radiatively induced)
$H\ell_i\ell_j$ LFV couplings are unavoidable \cite{bkl}.
These effects have been widely discussed in the recent literature
through the study of several processes, namely
$\tau\!\rightarrow\!\ell_j\ell_k\ell_k$ \cite{bkl}, $\tau\!\rightarrow\!\mu\eta$
\cite{sher}, $\mu-e$ conversion in nuclei \cite{kitano},
$B\!\rightarrow\!\ell_j\tau$ \cite{ellis},
$H\!\rightarrow\!\ell_j\ell_k$ \cite{andrea} and $\ell_i\!\rightarrow\!\ell_j\gamma$
\cite{hmio}.\\ In this Letter we analyze the LFV decay channels
of the purely leptonic $\pi^{\pm}$ and $K^{\pm}$ decays and discuss a
possible way to detect LFV SUSY effects through a deviation from the 
$\mu-e$ universality. One could naively think that SUSY effects in the LFV 
channels  $M\rightarrow \ell_i\nu_k$ are further suppressed with respect to 
the LFC ones. On the contrary, we show that charged Higgs mediated SUSY LFV 
contributions, in particular in the kaon decays into an electron or a muon 
and a tau neutrino, can be strongly enhanced.\\
The quantity which now accounts for the deviation from the $\mu-e$
universality reads:
$$
R^{LFV}_{\pi,K}=
\frac{\sum_i\Gamma(\pi(K)\rightarrow e\nu_i)}
{\sum_i\Gamma(\pi(K)\rightarrow \mu\nu_{i})}
\,\,\,\,\,\,\,\,\,\,\,\,\,\,\,\,\,\,\,\,\,\,\,\,\,\,\,\,i= e,\mu,\tau.
$$
with the sum  extended over all (anti)neutrino flavors 
(experimentally one determines only the charged lepton flavor in the
decay products).

\begin{figure}[t]
\includegraphics[scale=0.80]{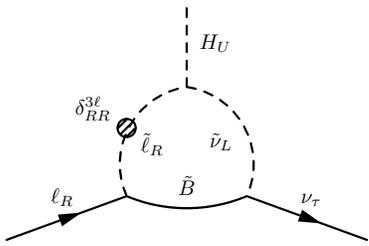}
\hskip 0.4 cm
\caption{\label{fig} 
Contribution to the effective  $\bar{\nu}_{\tau} \ell_R H^+$ coupling.}
\end{figure}

The dominant SUSY contributions to $R^{LFV}_{\pi,K}$ arise from the
charged Higgs exchange. The effective LFV Yukawa couplings we consider are
(see Fig.~\ref{fig}):
\beq
\label{coupl1}
\ell H^{\pm}{\nu_{\tau}}\rightarrow
\frac{g_2}{\sqrt2}\frac{{m_{\tau}}}{M_W}\Delta^{3l}_{R}\tan^{2}\!\beta
\,\,\,\,\,\,\,\,\,\,\,\,\ell=e,\mu.
\eeq
Crucial to our result is the quadratic dependence on $\tan\!\beta$ in the above coupling:
one power of $\tan\!\beta$ comes from the trilinear scalar coupling in Fig.1, while 
the second one is a specific feature of the above HRS mechanism.

The $\Delta^{3\ell}_{R}$ terms are induced at one loop level by
the exchange of Bino (see Fig.~\ref{fig}) or Bino-Higgsino and sleptons.
Since the Yukawa operator is of dimension four, the quantities
$\Delta^{3\ell}_{R}$ depend only on ratios of SUSY masses, 
hence avoiding SUSY decoupling. In the so called MI approximation 
the expression of $\Delta^{3\ell}_{R}$ is given by:
\beq
\Delta^{3\ell}_{R}\!\simeq\! \frac{\alpha_{1}}{4\pi}\mu M_1 m^{2}_{R} \delta^{3\ell}_{RR}
\left[I^{'}\!(M^{2}_{1},\mu^2,m^{2}_{R})\!-\!(\mu\!\leftrightarrow\! m_{L})
\right]
\eeq
where $\mu$ is the the Higgs mixing parameter, $M_1$ is the Bino ($\tilde{B}$) mass
and $m^{2}_{L(R)}$ stands for the left-left (right-right) slepton mass matrix entry.
The LFV MIs, i.e. 
$\delta^{3\ell}_{XX}\!=\!({\tilde m}^2_{\ell})^{3\ell}_{XX}/m^{2}_{X}$ $(X=L,R)$,
are the off-diagonal flavor changing entries of the slepton mass matrix.
The loop function $I^{'}(x,y,z)$ is such that $I^{'}(x,y,z)= dI(x,y,z)/d z$, 
where $I(x,y,z)$ refers to the standard three point one-loop integral which 
has mass dimension -2.
Following the thorough analysis in \cite{andrea}, it turns out that
$\Delta^{3\ell}_{R}\lesssim 10^{-3}$.

Making use of the LFV Yukawa coupling in Eq.~(\ref{coupl1}),
it turns out that the dominant contribution to $\Delta r^{e-\mu}_{NP}$ reads:
\beq
\label{lfv}
R^{LFV}_{K}\simeq R^{SM}_{K}
\left[1+\left(\frac{m^{4}_{K}}{M^{4}_{H}}\right)
\!\left(\frac{m^{2}_{\tau}}{m^{2}_{e}}\right)|\Delta^{31}_{R}|^2\,
\tan^{\!6}\!\beta\right].
\eeq
In Eq.~(\ref{lfv}) terms proportional to $\Delta^{32}_{R}$ are neglected given that
they are suppressed by a factor $m^{2}_{e}/m^{2}_{\mu}$ with respect to the term
proportional to $\Delta^{31}_{R}$.
Taking $\Delta^{31}_{R}\!\simeq\!5\cdot 10^{-4}$ accordingly to what said above,
$\tan\beta\!=\!40$ and $M_{H}\!=\!500 GeV$ we end up with
$R^{LFV}_{K}\!\simeq\!R^{SM}_{K}(1+0.013)$.
We see that in the large (but not extreme) $\rm\tan\beta$ regime and with
a relatively heavy $H^{\pm}$, it is possible to reach contributions
to $\Delta r^{e-\mu}_{\!K\,SUSY}$ at the percent level thanks to the
possible LFV enhancements arising in SUSY models.

Turning to pion physics, one could wonder whether the analogous quantity 
$\Delta r^{e-\mu}_{\!\pi\,SUSY}$ is able to constrain SUSY LFV.
However,  the correlation between
$\Delta r^{e-\mu}_{\!\pi\,SUSY}$ and $\Delta r^{e-\mu}_{\!K\,SUSY}$:
\beq
\label{lfvpi}
\Delta r^{e-\mu}_{\pi\,SUSY}\simeq\left(\frac{m_d}{m_u+m_d}\right)^{2}
\left(\frac{m^{4}_{\pi}}{m^{4}_{k}}\right)
\Delta r^{e-\mu}_{\!K\,SUSY}
\eeq
clearly shows that the constraints on $\Delta r^{e-\mu}_{\!K\,susy}$ force
$\Delta r^{e-\mu}_{\pi\,susy}$ to be much below its actual exp. upper bound.

Obviously, a legitimate worry when witnessing such a huge SUSY contribution 
through LFV terms is whether the bounds on LFV  tau decays, like
$\tau\rightarrow eX$ (with $X=\gamma,\eta,\mu\mu$), are respected \cite{hmio}.
Higgs mediated $Br(\tau\rightarrow \ell_j X)$ and $\Delta r^{e-\mu}_{\!K\,Susy}$
have exactly the same SUSY dependence; hence, we can compute the upper bounds of
the relevant LFV tau decays which are obtained for those values of the SUSY parameters
yielding $\Delta r^{e-\mu}_{\!K\,SUSY}$ at the percent level.
We obtain $Br(\tau\rightarrow eX)\leq 10^{-10}$ \cite{hmio}.
Given the exp. upper bounds on the LFV $\tau$ lepton decays \cite{belletmg},
we conclude that it is possible to saturate the upper bound on
$\Delta r^{e-\mu}_{\!K\,Susy}$ while remaining much below the present and expected
future upper bounds on such LFV decays. There exist other SUSY contributions to LFV
$\tau$ decays, like the one-loop neutralino-charged slepton exchanges,
for instance, where there is a direct dependence on the quantities $\delta^{3j}_{RR}$.
Given that the existing bounds on the leptonic $\delta_{RR}$ involving transitions
to the third generation are rather loose \cite{1mio}, it turns out that also these 
contributions fail to reach the level of exp. sensitivity for LFV $\tau$ decays.

\section{On the sign of $\Delta r^{e-\mu}_{\!SUSY}$}
%\textbf{On the sign of $\Delta r^{e-\mu}_{\!SUSY}$:}
The above SUSY dominant contribution to $\Delta r^{e-\mu}_{\!NP}$
arises from LFV channels in the $K\!\rightarrow\!e\nu$ mode,
hence without any interference effect with the SM contribution.
Thus, it can only increase the value of $R_{K}$ with respect to the SM
expectation. On the other hand, the recent NA48/2 result exhibits a
central value lower than $R_{K}^{SM}$
(and, indeed, also lower than the previous PDG central value).
One may wonder whether SUSY could account for such a lower $R_{K}$.
Obviously, the only way it can is through terms which, contributing to
the LFC $K\!\rightarrow\!l\nu_{l}$ channels, can interfere (destructively)
with the SM contribution. We already commented that SUSY LFC contributions
are subdominant. However, one can envisage the possibility of making use
of the large LFV contributions to give rise to LFC ones through double
LFV MI that, as a final effect, preserves the flavour.

To see this point explicitly, we derive the corrections to the LFC 
$H^{\pm}\ell\nu_{\ell}$ vertices induced by LFV effects
\beq
\label{coupl2}
\ell H^{\pm}{\nu_{\ell}}\!\rightarrow\! 
\frac{g_2}{\sqrt 2}\frac{{m_{\ell}}}{M_W}\tan\!\beta
\left(\!1\!+\!\frac{m_{\tau}}{m_{\ell}}\Delta^{\ell\ell}_{RL}
\tan\!\beta\!\right)\,,
%\,\,\,,l=e,\mu
\eeq
where $\Delta^{\ell\ell}_{RL}$ is generated by the same diagram 
as in Fig.~\ref{fig} but with an additional $\delta^{3\ell}_{LL}$ MI
in the sneutrino propagator.
In the MI approximation, $\Delta^{\ell\ell}_{RL}$ is given by
\beq
\Delta^{\ell\ell}_{RL}\!\simeq\! 
-\frac{\alpha_{1}}{4\pi}\mu M_1 m^{2}_{L} m^{2}_{R}
\,\delta^{\ell 3}_{RR}\delta^{3\ell}_{LL}\,I^{''}\!(M^{2}_{1},m^{2}_{L},m^{2}_{R})\,,
\eeq
where $I^{''}(x,y,z)= d^2I(x,y,z)/dydz$.
In the large slepton mixing case, $\Delta^{\ell\ell}_{RL}$ terms are of the
same order of $\Delta^{3\ell}_{R}$
\footnote{Im($\delta^{13}_{RR}\delta^{31}_{LL}$)
is strongly constrained by the electron electric dipole moment \cite{masina}.
However, sizable contributions to $R^{LFV}_{K}$ can still be induced by Re($\delta^{13}_{RR}\delta^{31}_{LL}$).}.
These new effects modify the previous $R^{LFV}_{K}$ expression in the following way:
\bea
\label{lfclfv}
R^{LFV}_{K}\simeq R^{SM}_{K}
&\bigg[&\bigg|1\!-\!\frac{m^{2}_{K}}{M^{2}_{H}}
\frac{m_{\tau}}{m_{e}}\Delta^{11}_{RL}\,\tan^{\!3}\!\beta\bigg|^{2} +
\nonumber \\
&+&\bigg(\frac{m^{4}_{K}}{M^{4}_{H}}\bigg)
\!\bigg(\frac{m^{2}_{\tau}}{m^{2}_{e}}\bigg)
|\Delta^{31}_{R}|^2\,\tan^{\!6}\!\beta
\bigg].
\eea
In the above expression, besides the contributions reported in Eq.~(\ref{lfv}),
we included also the interference between SM and SUSY LFC terms
(arising from a double LFV source).
Setting the parameters as in the example of the above section and if  
$\Delta^{11}_{RL}\!=\!10^{-4}$ we get
$R^{LFV}_{K}\!\simeq\! R^{SM}_{K}(1-0.032)$,
that is just within the expected exp. resolution 
reachable by NA48/2 once all the available data will be analyzed.
Finally, we remark that the above effects do not spoil the pion physics 
constraints.

\section{Conclusions}
%\textbf{Conclusions:}
Our Letter shows that, rather surprisingly, a precise measurement 
of the flavor conserving $K_{\ell2}$ decays may shed light on the size 
of LFV in new physics. Since neutrino masses and oscillations clearly 
point out that lepton flavor number is violated and since new physics 
(for instance supersymmetric versions of models with see-saw mechanism 
for neutrino masses \cite{masierorew}) is known to have the potentiality for 
(large) enhancements of such LFV with respect to the SM, 
we emphasize the importance of further improving the exp. 
sensitivity on $R_K$ as a particularly interesting probe of such 
new physics effects.

\textit{Acknowledgements:}
We thank G. Isidori for valuable discussions and N.Tantalo
for his collaboration at the early stage of this work.
We acknowledge support from the RTN European project MRTN-CT-2004-503369.

\end{document}